**Interlayer dark excitons in a van der Waals heterostructure**

Rundong Ma[1], Konstantin Davydov[1], Liuxin Gu[1], Lifu Zhang[1], Hassan Alnatah[1], Beini Gao[1], Ruihao Ni[1], Suji Park[2], Houk Jang[2], Takashi Taniguchi[3], Kenji Watanabe[4], You Zhou[1,5,†]

[1]Department of Materials Science and Engineering, University of Maryland, College Park, MD 20742, USA

[2]Center for Functional Nanomaterials, Brookhaven National Laboratory, Upton, NY 11973, USA

[3]Research Center for Electronic and Optical Materials, National Institute for Materials Science, 1-1 Namiki, Tsukuba 305-0044, Japan

[4]Research Center for Materials Nanoarchitectonics, National Institute for Materials Science, 1-1 Namiki, Tsukuba 305-0044, Japan

[5]Maryland Quantum Materials Center, College Park, MD 20742, USA

†To whom correspondence should be addressed: youzhou@umd.edu

**Abstract**

Interlayer excitons in transition metal dichalcogenide (TMD) heterostructures exhibit long lifetimes and long-range transport, making them promising for excitonic devices and quantum many-body phases, such as Bose–Einstein condensates. Achieving these goals requires a precise understanding of spin-allowed bright and nominally spin-forbidden dark excitons, because the lowest-energy exciton species governs population, transport, and condensation. Despite substantial progress, unambiguously distinguishing singlet and triplet interlayer excitons has been challenging, as moiré excitons in these heterostructures can mimic their optical signatures. Here, we report the direct spectroscopic identification of bright (singlet) and dark (triplet) interlayer excitons in high-quality, dual-gated $WSe_2$/hBN/$WSe_2$ homobilayers. Electric-field-dependent photoluminescence and reflectance reveal two momentum-direct interlayer transitions with distinct spin configurations. The interlayer dark excitons obey selection rules that differ from those of bright excitons. Strikingly, interlayer dark excitons retain strong valley polarization, even with their ultralong lifetime exceeding microseconds. Finally, we demonstrate twist-angle control, wherein twist-induced electron–hole momentum mismatch modulates interlayer exciton emission. These results provide critical insights into the electronic and excitonic structure of TMD heterostructures, opening new avenues for excitonic many-body physics and optoelectronic devices.

## Introduction

Interlayer excitons (IXs) in TMD van der Waals heterostructures have emerged as a versatile platform for exploring light–matter interactions and many-body quantum physics[1–5]. They are spatially indirect, with electron and hole wavefunctions localized in separate layers. Such layer separation generates an out-of-plane electric dipole and increases radiative lifetimes by orders of magnitude relative to intralayer excitons[6–9]. The dipole enables Stark tuning of the IX energy with vertical electric fields, while the long lifetimes support micrometer-scale transport and diffusion[10,11]. Together, these features make TMD heterostructures an ideal system to study exciton dynamics and many-body phenomena—including transport[10,12], cooling[13,14], trapping[15–18], as well as Bose–Einstein condensation and superfluidity at elevated temperatures[1–4,19,20].

A defining property of excitons is their spin configuration: the electron and hole spins can either be antiparallel (singlet, nominally "bright") or parallel (triplet, nominally "dark"). In TMDs, strong spin–orbit coupling and valley degrees of freedom further differentiate these states, producing distinct spin–valley physics and optical selection rules[21–23]. Although intralayer bright and dark excitons are well-characterized in monolayer TMDs, distinguishing the singlet and triplet IXs is far more challenging. On the one hand, broken mirror symmetry can endow nominally spin-forbidden IXs with a finite optical dipole, whose selection rule sensitively depends on local atomic stacking[24–29], so that brightness and helicity alone cannot distinguish singlets from triplets. On the other hand, moiré potentials in TMD heterostructures generate quantum-confined excitons whose selection rules resemble those of singlet and triplet IXs[30–32], further obscuring their assignment. Elucidating the spin character of IXs thus remains an outstanding challenge, with critical implications for excitonics, valleytronics, and exciton condensates, as the lowest-energy spin configuration governs the exciton population, dynamics, and collective phases such as Bose–Einstein condensation.

## Results

Here, we investigate spin states of IXs in a dual-gated moiré-free bilayer system, composed of two $WSe_2$ monolayers separated by an atomically thin hBN spacer (**Fig. 1a**; optical images of devices D1–D4 are shown in **Fig. S1**). The hBN layer decouples the two TMDs, which suppresses the moiré potential and preserves the direct band gap at the K and K′ valleys, while still permitting weak interlayer tunneling. This tunneling allows electrons and holes in separate layers to recombine directly, producing IX emission, as previously observed in $MoSe_2$/hBN/$MoSe_2$ heterostructures[11,33,34]. In $WSe_2$, in contrast to $MoSe_2$, the dark excitons lie below the bright excitons[35–37], potentially allowing spin-singlet and spin-triplet states to be probed via photoluminescence (PL).

**Figure 1b** shows the electric-field-dependent PL spectrum of a nearly 0°-aligned device (D1) at 6K. Interlayer excitons with a linear Stark shift can be clearly distinguished from intralayer ones, whose energies remain unchanged. The PL map is symmetric in field, indicating two degenerate IX branches with opposite dipole orientations that split under a finite field. Examining the IX emission closely, we find two species, labeled $IX_B$ and $IX_D$, with a similar Stark shift slope but different energies, suggesting the same layer separation but distinct origins.

To track how these features evolve with field, we present spectra at three representative fields (**Fig. 1c**). At a low electric field, the emission resembles that of a monolayer, with dominant peaks from neutral bright excitons at ~1.73 eV ($X_0$) and charged excitons ($X_S^-$ and $X_t^-$), arising from residual electrons in $WSe_2$. The $X_0$ emissions from the two monolayers are nearly degenerate, indicating minimal strain variation between them. We also observe the spin-forbidden dark exciton $X_D$ ~43 meV below $X_0$, together with its phonon replicas, charged, and brightened counterparts at lower energies[37–40].

With increasing electric field *E*, we observe enhanced $X_S^-$ and $X_t^-$ emission and reduced $X_0$ emission, due to electron accumulation in one layer under the polarizing field (**Fig. 1b**). At *E*~0.1 V/nm, $IX_B$ and $X_0$ become degenerate, enhancing $IX_B$ emission and suppressing $X_0$ (**Figs. 1b and 1c**). At ~0.12 V/nm, when $IX_D$ approaches $X_D^-$, we observe strong enhancement of $IX_D$ and intralayer dark excitons ($X_D^-$ and $X_{DB}^-$), while bright exciton emission is suppressed. Above this field, all intralayer dark excitons are quenched. The emission is dominated by $IX_D$, while $X_0$ recovers to its zero-field level.

We attribute $IX_B$ and $IX_D$ to momentum-direct (K–K) interlayer excitons, with different spin configurations: $IX_B$ is a spin-singlet (bright), $IX_D$ a spin-triplet (dark), which lies lower in energy due to spin–orbit coupling in the conduction band (**Fig. 1d**). At zero field, IXs with weaker oscillator strength and higher energy than intralayer states have negligible emission. As the Stark effect shifts IX energies toward the intralayer excitons, finite interlayer tunneling hybridizes spin-matched states, $X_0$ with $IX_B$ and $X_D$ with $IX_D$. As $IX_B$ approaches $X_0$, it borrows oscillator strength and brightens, whereas $IX_D$ remains dark. At higher fields, $IX_D$ becomes resonant with and then falls below $X_D$, which funnels $X_D$ population into $IX_D$ and quenches other dark-exciton lines. $IX_D$ intensity decreases with increasing temperature (**Fig. S2**), as expected for the lowest-lying spin-forbidden state and consistent with intralayer dark excitons[35,38,41].

With the assignment established, we can now quantify the properties of IXs and the underlying band structure. From the slope of the Stark effect in **Fig. 1b**, we estimate the out-of-plane electric dipole moment of $IX_D$ and $IX_B$ both to be $\sim e * 0.98 \pm 0.01$ nm, in good agreement with the expected interlayer separation of two-layer hBN spacer plus the finite thickness of $WSe_2$ (see Supplementary Discussion I for dipole moment estimation). We further estimate the $IX_B$ binding energy by subtracting the zero-field $IX_B$ energy from the quasiparticle bandgap (see Supplementary Discussion II), yielding an $IX_B$ binding energy of ~73 meV, consistent with theoretical calculations[19,42] (see Table S1 for its dependence on hBN thickness).

The measured $IX_B$–$IX_D$ splitting, ~19 meV, provides critical insight into the $WSe_2$ band structure. This bright–dark splitting arises from a combination of conduction band spin–orbit coupling $\Delta_C$, differences in exciton Coulomb binding energy, and exchange interaction. The exchange energy difference is expected to be small, given the spatial separation of electron and hole wavefunctions. Since the upper conduction band has a lighter mass, $IX_B$ would bind more weakly than $IX_D$, implying the $IX_B$–$IX_D$ splitting exceeds $\Delta_C$. Thus, the observed ~19 meV serves as an upper bound on $\Delta_C$, consistent with the previously reported values[36,39,43] and theoretical calculations[44,45].

To further corroborate the assigned spin configuration, we perform field-dependent reflectance measurements and observe finite $IX_B$ absorption near the $IX_B$–$X_0$ crossover (**Fig. 2a**). By contrast, no absorption appears below 1.70 eV, where $IX_D$ and dark excitons are expected (**Fig. S3a**). These observations are consistent with hybridization that transfers oscillator strength only between same-spin pairs: $IX_B$ borrows from bright $X_0$, whereas $IX_D$ gains little from spin-forbidden $X_D$ and remains undetectable.

To probe the interactions between carriers and IXs, we measure PL under an asymmetric gating scheme, $V_{BG} = \alpha V_{TG} + \delta$, with $\delta$ setting a strong out-of-plane electric field. This field polarizes the doped carriers in one layer while leaving the other intrinsic, as evidenced by intralayer exciton emission. As the Fermi level reaches the band edge in either layer, we observe a lower-energy peak below $IX_D$, indicating charged $IX_D$ with an exciton–carrier binding energy of a few meV (**Fig. 2b**). In the more heavily doped regimes ($V_{TG} < -5.8$ V and $V_{TG} > -2$ V), the charged $IX_D$ blueshifts, primarily because the highly doped layer screens and reduces the net field between the two TMDs[11] (see Supplementary Discussion III).

The distinct spin configurations of $IX_B$ and $IX_D$ entail dramatically different relaxation dynamics, as revealed by time-resolved PL (**Figs. 2c** and **S3b**). In D1, the observed $IX_B$ decay is limited by the instrument response function, setting an upper bound on its lifetime of ~0.3 ns. By contrast, $IX_D$ exhibits a lifetime of hundreds of nanoseconds, which further increases with the electric field (**Fig. S3b**). With a thicker hBN spacer that further reduces tunneling, the $IX_D$ lifetime extends to several microseconds (D2 in **Fig. 2c**), orders of magnitude longer than in heterobilayers without a spacer[6,10].

We further study the valley dynamics of $IX_B$ and $IX_D$ using circularly polarized 1.95 eV excitation and analyzing the resulting PL polarization. Intriguingly, $IX_B$ and $IX_D$ emit in opposite circular helicities (**Figs. 3a, 3b and S4**). We quantify this with the degree of circular polarization (DOCP), defined as $DOCP = \frac{I_{co} - I_{cross}}{I_{co} + I_{cross}}$, where $I_{co}$ and $I_{cross}$ denote the PL intensities in the co- and cross-polarized detection channels, respectively (**Fig. 3c**). The DOCP for both $IX_B$ and $IX_D$ reaches approximately +40% and −40%, with magnitudes comparable to those of intralayer excitons. The surprising persistence of strong IXD polarization despite its ~µs lifetime indicates slow valley depolarization and that the polarization is determined primarily by the initial excitation and intralayer relaxation (**Fig. S5**).

These contrasting helicities suggest distinct selection rules for $IX_B$ and $IX_D$ (**Fig. 3d**). In $WSe_2$ monolayers with $C_3$ and mirror symmetry, the singlet (triplet) excitons couple with in-plane (out-of-plane) optical fields. In bilayers with broken mirror symmetry, such as under an external electric field, selection rules of IXs depend on both spin configuration and atomic registry, which is sensitive to the relative shift and twist between layers[29]. Therefore, selection rules alone cannot distinguish singlet/triplet excitons[26,46–48] from moiré excitons[30–32]. By suppressing moiré effects, our platform isolates the underlying singlet–triplet physics.

Nevertheless, the measured selection rules and oscillator strengths are unexpected for our nominally 0°-aligned (R-type) devices. In particular, for $R_X^X$ stacking, $IX_B$ shares the selection rules of $X_0$, whereas $IX_D$ couples to the out-of-plane field. While experimentally $IX_B$ follows this expectation, $IX_D$ does not, as further confirmed by Fourier-plane imaging (**Fig. S6**). Instead, the measured $IX_D$ selection rule is more consistent with a different stacking order, $R_X^M$. Moreover, while theory anticipates comparable oscillator strengths for $IX_B$ and $IX_D$, their markedly different absorption and lifetimes indicate otherwise. These discrepancies point to physics not captured by current theoretical models. One possibility is that a small interlayer twist creates domains with distinct local registries, each with different selection rules, energies, and lifetimes, such that recombination is dominated by specific registries. A complete account will require modeling IX dynamics in the presence of the hBN spacer, which we leave for future work.

Finally, we demonstrate the momentum control of the IXs by twisting the two TMD layers by ~1° in device D4. Unlike zero-twist devices, the field-dependent PL of D4 shows no IX emission (**Fig. 4a**). Nevertheless, the intralayer excitons quench sequentially with field: dark species near-completely disappear at ~0.14 V/nm (**Fig. 4b**), bright species only partially weaken, similar to **Fig. 1b**. This behavior arises from twist-induced momentum mismatch: the rotated Brillouin zones of the two $WSe_2$ layers (**Fig. 4c**) suppress interlayer tunneling and enhance IX lifetimes. Under an electric field, photoexcited intralayer excitons can either recombine radiatively or relax into lower-energy IX states. Dark excitons, with long radiative lifetimes, relax into IXs more readily and thus quench more strongly, whereas bright excitons, with picosecond lifetimes, are less affected. These trends imply that intralayer-to-IX relaxation in D4 is slower than the bright-exciton radiative lifetime (~ps) but faster than that of dark excitons (~100s of ps)[36,49], demonstrating the twist angle as an additional knob to tune tunneling and IX recombination.

## Conclusion

Our experiments provide direct spectroscopic identification of spin-allowed bright and spin-forbidden dark interlayer excitons in TMD heterostructures, yielding new insights into electronic band structure, interlayer exciton binding, relaxation, and their interactions with free carriers. The opposite helicities of the singlet and triplet states add an independent axis for valley encoding. Moreover, evidence for distinct recombination sites raises the intriguing prospect of selectively funneling singlet and triplet excitons into different atomic registries for nanoscale excitonic routing. The combination of long, electrically tunable lifetimes, robust valley polarization, and strong interactions provides a platform for exploring correlated exciton phases such as condensation and superfluidity and for storing and manipulating valley and spin degrees of freedom in excitonic and valleytronic devices.

## Methods

### Device design and fabrication

A schematic of the dual-gated $WSe_2$/hBN/$WSe_2$ device is shown in **Fig. 1a** of the main text. The top and bottom hBN gate dielectrics are approximately 30 nm thick. The hBN spacer between the two $WSe_2$ monolayers has a typical thickness of 0.5–1 nm. The constituent layers, graphite and hBN flakes of varying thicknesses, were mechanically exfoliated from bulk crystals onto Si substrates with a 285 nm $SiO_2$ layer. Monolayer $WSe_2$ was supplied by the Quantum Material Press (QPress) facility at the Center for Functional Nanomaterials (CFN), Brookhaven National Laboratory (BNL). The device was fabricated using a dry transfer technique on a transfer station built by Everbeing Int'l Corp. The metal electrodes, containing 5 nm Cr and 80 nm Au, were deposited using electron-beam lithography and thermal evaporation.

### Optical spectroscopy

All optical spectroscopy measurements were performed using our home-built confocal microscope equipped with an Attocube cryostat (Attocube 800). An apochromatic objective with a numerical aperture (NA) of 0.82 was used. PL measurements were conducted using a continuous-wave (CW) 635 nm diode laser as the excitation source. Reflectance measurements were performed using a broadband halogen lamp (Thorlabs), and time-resolved PL measurements were carried out using a YSL SC-Pro-M supercontinuum laser. The diode laser has a diffraction-limited spot size of around 1 μm, and its power can be adjusted continuously. The reflectance data were normalized by the reflectance from the nearby sample region without TMDs. All the spectra were collected using a Horiba iHR320 spectrometer equipped with a 300 grooves/mm grating and a Synapse-Plus back-illuminated deep-depletion CCD camera. All data were obtained at 6 K unless otherwise noted, with no magnetic field.

**Acknowledgements**

We acknowledge support from the Army Research Office W911NF2510066 and the National Science Foundation DMR-2145712. Sample fabrication was supported by the U.S. Department of Energy, Office of Science, Office of Basic Energy Sciences under Award No. DE-SC-0022885. This research used the Quantum Material Press (QPress) of the Center for Functional Nanomaterials (CFN), which is a U.S. Department of Energy Office of Science User Facility, at Brookhaven National Laboratory under Contract No. DE-SC0012704. K.W. and T.T. acknowledge support from the JSPS KAKENHI (Grant Numbers 20H00354, 21H05233, and 23H02052) and the World Premier International Research Center Initiative (WPI), MEXT, Japan, for hBN synthesis.

**Author contributions**

Y.Z. and R.M. conceived the project. R.M. fabricated the samples and performed the experiments. K.D., L.G., R.N., L.Z., S.P., and H.J. assisted with sample fabrication. K.D., L.G., B.G. and H.A. helped with optical measurements. R.M., and Y.Z. contributed to the data analysis and theoretical understanding. T.T. and K.W. provided hexagonal boron nitride samples. R.M. and Y.Z. wrote the manuscript with extensive input from the other authors.

**Competing interests**

The authors declare no competing interests.

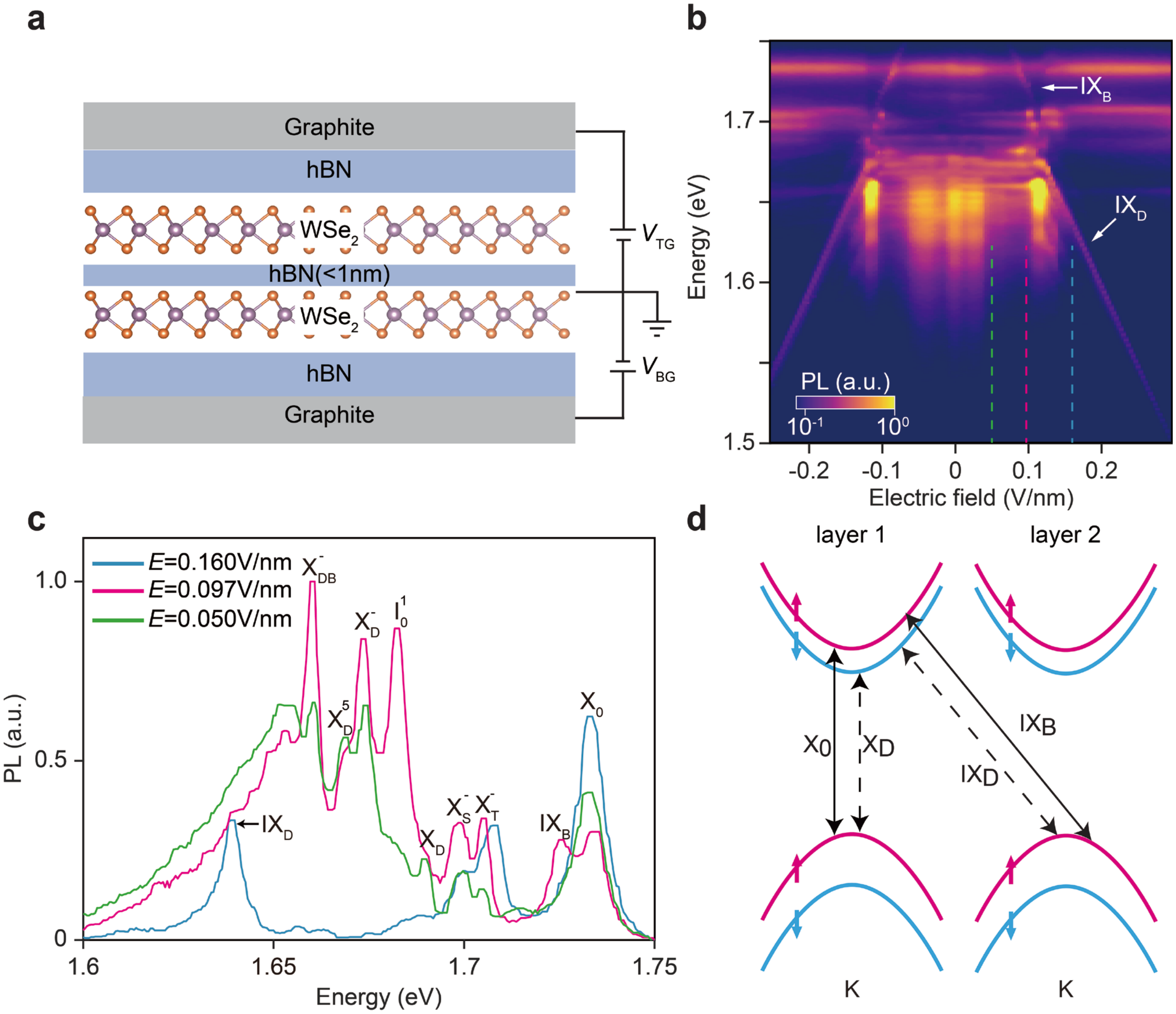

**Fig. 1 | Electrical control of interlayer excitons.** (a) Schematic of a dual-gate device. (b) Electric-field-dependent PL spectra when both layers are intrinsic. Dashed lines indicate where the spectra in (c) are collected. (c) Representative PL spectra at 0.050 V/nm, 0.097 V/nm, and 0.160 V/nm. The spectra show the neutral intralayer exciton ($X_0$), the intralayer dark exciton ($X_D$) and its phonon replicas ($X_D^5$), indirect exciton phonon replicas ($I_0^1$), the charged dark exciton ($X_D^-$), and the brightened dark exciton ($X_{DB}^-$), located approximately 43, 58, 51, 63, and 72 meV below $X_0$, respectively. At 0.097 V/nm, enhanced emission is observed from interlayer bright exciton ($IX_B$), $I_0^1$, $X_D^-$ and $X_{DB}^-$. At 0.160 V/nm, emission from the interlayer dark exciton ($IX_D$) dominates. (d) Layer-resolved band diagram, showing $X_0$, $X_D$, the interlayer bright exciton ($IX_B$), and the interlayer dark exciton ($IX_D$).

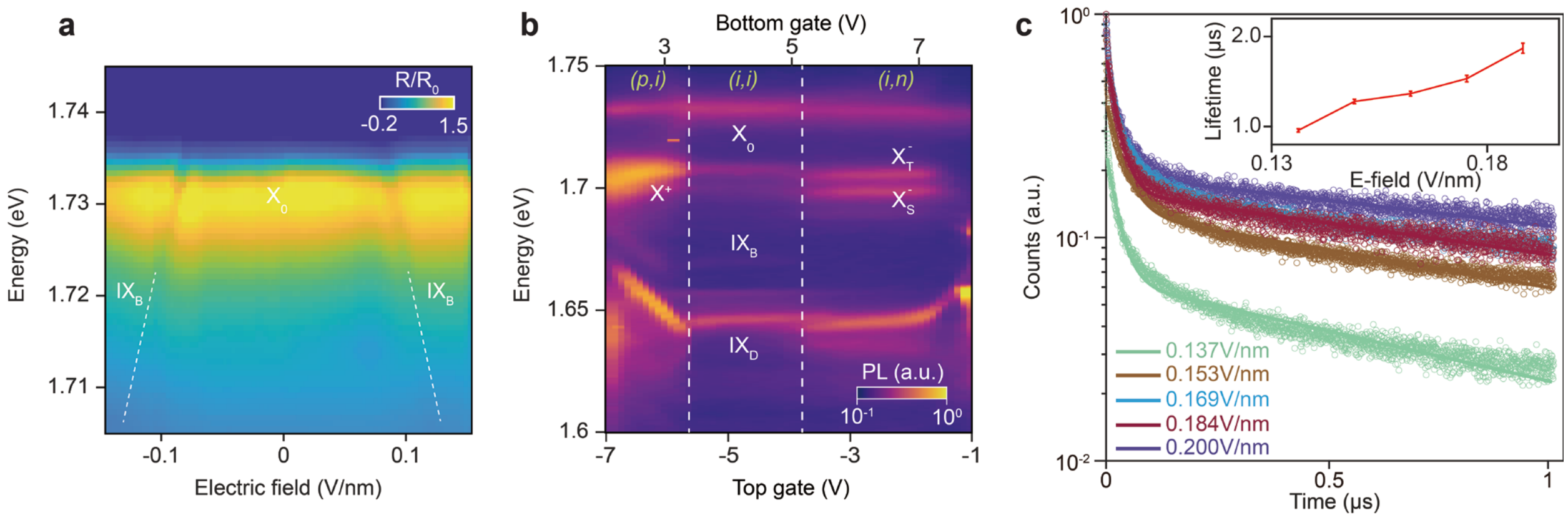

**Fig. 2 | Reflectance, doping-dependence, and lifetime dynamics of the interlayer excitons.** (a) Electric field dependence of reflectance spectra. Dashed lines trace $IX_B$. (b) Doping dependence of $IX_D$ PL spectra under an electric field ($V_{BG}$ = 0.96 $V_{TG}$ + 8.8 V). (c) Time-resolved photoluminescence (TRPL) measurement of $IX_D$ in D2 under different electric fields when both layers are intrinsic. The inset shows the fitted lifetime as a function of the applied electric field.

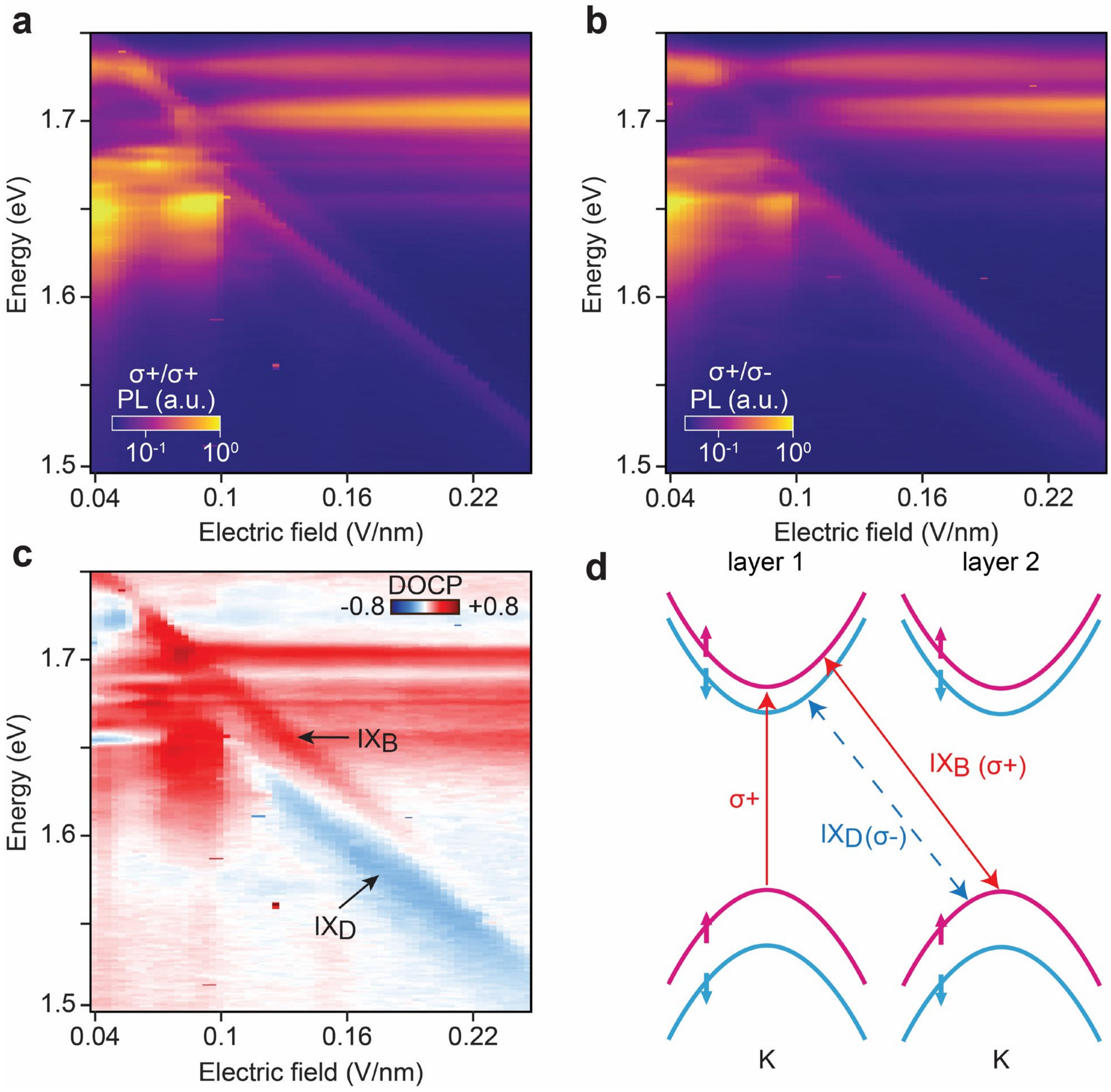

**Fig. 3 | Valley polarization of interlayer excitons.** (a, b) Electric-field-dependent PL under (**a**) co-circular and (b) cross-circular excitation and detection. (c) Degree of circular polarization (DOCP) as a function of the electric field. (d) Schematic of selection rules of $IX_B$ and $IX_D$.

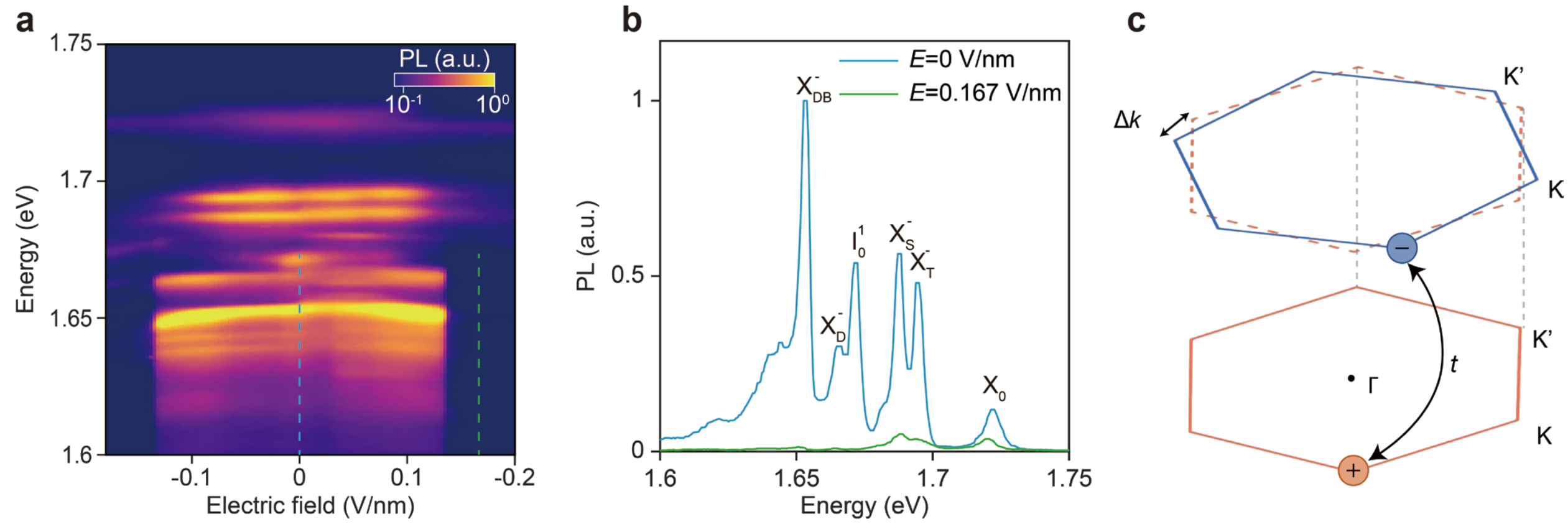

**Fig. 4 | Electrical control of excitons in a twisted $WSe_2$/hBN/$WSe_2$ heterostructure.** (a) Electric-field-dependent PL with both layers intrinsic. Dashed lines indicate where the spectra in (b) are collected. (b) Representative PL spectra at 0 V/nm (blue) and 0.167 V/nm (green). (c) Schematic of twist-induced momentum mismatch that suppresses interlayer tunneling and IX emission.

## References

1. Wang, Z. *et al.* Evidence of high-temperature exciton condensation in two-dimensional atomic double layers. *Nature* **574**, 76–80 (2019).
2. Xie, M. & MacDonald, A. H. Electrical reservoirs for bilayer excitons. *Phys. Rev. Lett.* **121**, 067702 (2018).
3. Wu, F.-C., Xue, F. & MacDonald, A. H. Theory of two-dimensional spatially indirect equilibrium exciton condensates. *Phys. Rev. B* **92**, 165121 (2015).
4. Berman, O. L. & Kezerashvili, R. Y. Superfluidity of dipolar excitons in a transition metal dichalcogenide double layer. *Phys. Rev. B.* **96**, 094502 (2017).
5. Shi, Q. *et al.* Bilayer WSe2 as a natural platform for interlayer exciton condensates in the strong coupling limit. *Nat. Nanotechnol.* **17**, 577–582 (2022).
6. Rivera, P. *et al.* Observation of long-lived interlayer excitons in monolayer MoSe2-WSe2 heterostructures. *Nat. Commun.* **6**, 6242 (2015).
7. Rivera, P. *et al.* Valley-polarized exciton dynamics in a 2D semiconductor heterostructure. *Science* **351**, 688–691 (2016).
8. Fang, H. *et al.* Strong interlayer coupling in van der Waals heterostructures built from single-layer chalcogenides. *Proc. Natl. Acad. Sci. U. S. A.* **111**, 6198–6202 (2014).
9. Jiang, Y., Chen, S., Zheng, W., Zheng, B. & Pan, A. Interlayer exciton formation, relaxation, and transport in TMD van der Waals heterostructures. *Light Sci. Appl.* **10**, 72 (2021).
10. Jauregui, L. A. *et al.* Electrical control of interlayer exciton dynamics in atomically thin heterostructures. *Science* **366**, 870–875 (2019).
11. Zhang, L. *et al.* Electrical control and transport of tightly bound interlayer excitons in a $MoSe_2$/hBN/$MoSe_2$ heterostructure. *Phys. Rev. Lett.* **132**, 216903 (2024).
12. Li, Z. *et al.* Interlayer exciton transport in MoSe2/WSe2 heterostructures. *ACS Nano* **15**, 1539–1547

(2021).

13. Policht, V. R. *et al.* Time-domain observation of interlayer exciton formation and thermalization in a MoSe2/WSe2 heterostructure. *Nat. Commun.* **14**, 7273 (2023).
14. Fang, H. *et al.* Localization and interaction of interlayer excitons in MoSe2/WSe2 heterobilayers. *Nat. Commun.* **14**, 6910 (2023).
15. Joe, A. Y. *et al.* Controlled interlayer exciton ionization in an electrostatic trap in atomically thin heterostructures. *Nat. Commun.* **15**, 6743 (2024).
16. Wang, W. & Ma, X. Strain-induced trapping of indirect excitons in $MoSe_2$/$WSe_2$ heterostructures. *ACS Photonics* **7**, 2460–2467 (2020).
17. Wu, Y.-C. *et al.* Highly tunable valley polarization of potential-trapped moiré excitons in WSe2/WS2 heterojunctions. *Phys. Rev. Lett.* **134**, 256402 (2025).
18. Brotons-Gisbert, M. *et al.* Moiré-trapped interlayer trions in a charge-tunable WSe2/MoSe2 heterobilayer. *Phys. Rev. X.* **11**, 031033 (2021).
19. Fogler, M. M., Butov, L. V. & Novoselov, K. S. High-temperature superfluidity with indirect excitons in van der Waals heterostructures. *Nat. Commun.* **5**, 4555 (2014).
20. Cutshall, J. *et al.* Imaging interlayer exciton superfluidity in a 2D semiconductor heterostructure. *Sci. Adv.* **11**, eadr1772 (2025).
21. Xiao, D., Liu, G.-B., Feng, W., Xu, X. & Yao, W. Coupled spin and valley physics in monolayers of MoS2 and other group-VI dichalcogenides. *Phys. Rev. Lett.* **108**, 196802 (2012).
22. Mak, K. F., He, K., Shan, J. & Heinz, T. F. Control of valley polarization in monolayer MoS2 by optical helicity. *Nat. Nanotechnol.* **7**, 494–498 (2012).
23. Zeng, H., Dai, J., Yao, W., Xiao, D. & Cui, X. Valley polarization in MoS2 monolayers by optical pumping. *Nat. Nanotechnol.* **7**, 490–493 (2012).
24. Hsu, W.-T. *et al.* Negative circular polarization emissions from WSe2/MoSe2 commensurate heterobilayers. *Nat. Commun.* **9**, 1356 (2018).
25. Joe, A. Y. *et al.* Electrically controlled emission from singlet and triplet exciton species in atomically

thin light-emitting diodes. *Phys. Rev. B.* **103**, L161411 (2021).

26. Zhang, L. *et al.* Highly valley-polarized singlet and triplet interlayer excitons in van der Waals heterostructure. *Phys. Rev. B* **100**, 041402 (2019).
27. Wang, T. *et al.* Giant valley-Zeeman splitting from spin-singlet and spin-triplet interlayer excitons in WSe2/MoSe2 heterostructure. *Nano Lett.* **20**, 694–700 (2020).
28. Woźniak, T., Faria Junior, P. E., Seifert, G., Chaves, A. & Kunstmann, J. Exciton g factors of van der Waals heterostructures from first-principles calculations. *Phys. Rev. B.* **101**, 235408 (2020).
29. Yu, H., Liu, G.-B. & Yao, W. Brightened spin-triplet interlayer excitons and optical selection rules in van der Waals heterobilayers. *2D Mater.* **5**, 035021 (2018).
30. Tran, K. *et al.* Evidence for moiré excitons in van der Waals heterostructures. *Nature* **567**, 71–75 (2019).
31. Huang, L. *et al.* Moiré-orbital-resolved excitonic Mott insulating states and their optical and electric control in van der Waals heterostructures. *Phys. Rev. Lett.* **135**, 096902 (2025).
32. Seyler, K. L. *et al.* Signatures of moiré-trapped valley excitons in MoSe2/WSe2 heterobilayers. *Nature* **567**, 66–70 (2019).
33. Shimazaki, Y. *et al.* Strongly correlated electrons and hybrid excitons in a moiré heterostructure. *Nature* **580**, 472–477 (2020).
34. Schwartz, I. *et al.* Electrically tunable Feshbach resonances in twisted bilayer semiconductors. *Science* **374**, 336–340 (2021).
35. Zhang, X.-X., You, Y., Zhao, S. Y. F. & Heinz, T. F. Experimental evidence for dark excitons in monolayer $WSe_2$. *Phys. Rev. Lett.* **115**, 257403 (2015).
36. Zhang, X.-X. *et al.* Magnetic brightening and control of dark excitons in monolayer WSe2. *Nat. Nanotechnol.* **12**, 883–888 (2017).
37. Zhou, Y. *et al.* Probing dark excitons in atomically thin semiconductors via near-field coupling to surface plasmon polaritons. *Nat. Nanotechnol.* **12**, 856–860 (2017).
38. Liu, E. *et al.* Gate tunable dark trions in monolayer WSe2. *Phys. Rev. Lett.* **123**, 027401 (2019).

39. Yang, M. *et al.* Relaxation and darkening of excitonic complexes in electrostatically doped monolayer WSe2: Roles of exciton-electron and trion-electron interactions. *Phys. Rev. B.* **105**, 085302 (2022).

40. Liu, E. *et al.* Multipath optical recombination of intervalley dark excitons and trions in monolayer $WSe_2$. *Phys. Rev. Lett.* **124**, 196802 (2020).

41. Li, Z. *et al.* Emerging photoluminescence from the dark-exciton phonon replica in monolayer WSe2. *Nat. Commun.* **10**, 2469 (2019).

42. Van der Donck, M. & Peeters, F. M. Interlayer excitons in transition metal dichalcogenide heterostructures. *Phys. Rev. B.* **98**, 115104 (2018).

43. Kapuściński, P. *et al.* Rydberg series of dark excitons and the conduction band spin-orbit splitting in monolayer WSe2. *Commun. Phys.* **4**, 186 (2021).

44. Kormányos, A. *et al.* K · p theory for two-dimensional transition metal dichalcogenide semiconductors. *2d Mater.* **2**, 022001 (2015).

45. Echeverry, J. P., Urbaszek, B., Amand, T., Marie, X. & Gerber, I. C. Splitting between bright and dark excitons in transition metal dichalcogenide monolayers. *Phys. Rev. B* **93**, 121107 (2016).

46. Ciarrocchi, A. *et al.* Polarization switching and electrical control of interlayer excitons in two-dimensional van der Waals heterostructures. *Nat. Photonics* **13**, 131–136 (2019).

47. Zhao, Y. *et al.* Interlayer exciton complexes in bilayer MoS2. *Phys. Rev. B.* **105**, L041411 (2022).

48. Unuchek, D. *et al.* Valley-polarized exciton currents in a van der Waals heterostructure. *Nat. Nanotechnol.* **14**, 1104–1109 (2019).

49. Robert, C. *et al.* Fine structure and lifetime of dark excitons in transition metal dichalcogenide monolayers. *Phys. Rev. B.* **96**, 155423 (2017).